\documentclass[english]{revtex4}
\usepackage{graphicx}

\begin{document}

\title{Fermi matrix element with isospin breaking}

\author{P.A.M. Guichon}
\affiliation{SPhN-IRFU, CEA Saclay, F91191 Gif sur Yvette, France}
\author{A.W. Thomas}
\affiliation{CSSM, School of Chemistry and Physics,University of Adelaide, SA 5005 Australia}
\author{K.~Saito}
\affiliation{Department of Physics, Faculty of Science and Technology, Tokyo University of Science,
2641, Yamazaki, Noda, Chiba, 278-8510, Japan}
%
\begin{abstract}
Prompted by the level of accuracy now being achieved in tests of the
unitarity of the CKM matrix, we consider the possible
modification of the Fermi matrix element for
the $\beta$-decay of a neutron, including possible in-medium and isospin
violating corrections. While the nuclear modifications lead to very
small corrections once the Behrends-Sirlin-Ademollo-Gatto theorem is
respected, the
effect of the $u-d$ mass difference on the conclusion concerning $V_{ud}$
is no longer insignificant. Indeed, we suggest that the correction to the
value of $|V_{ud}|^2 \, + \, |V_{us}|^2 \, + \, |V_{ub}|^2$ is at the
level of $10^{-4}$.
\end{abstract}
\maketitle
\section{Introduction}
The unitarity of the Cabbibo-Kobayashi-Maskawa
(CKM) matrix, ${\cal U}$, is a
fundamental aspect of the Standard Model and any deviation would
represent clear evidence of new physics. The most accurate test of
unitarity involves the (11) element of
${\cal U}^\dagger
{\cal U}$~\cite{Wilkinson:1994ua,Wilkinson:1990qu,Towner:1995za,Towner:2010zz},
namely
\begin{equation}
|V_{ud}|^2 \, + \, |V_{us}|^2 \, + \, |V_{ub}|^2 \, = \, 0.99995 \pm
0.00061 \, .
\label{eq:unitarity}
\end{equation}
This remarkable result is primarily a triumph of decades of precise
studies of super-allowed nuclear $\beta$-decay, which at the present
time yields a more accurate value of $V_{ud}$, namely $V_{ud} =
0.97425 \pm 0.00022$, than studies of the free neutron.

In the light of the remarkable level of accuracy that has been obtained,
as well as the improvements anticipated in the near future, we feel that
it is timely to address once more the issue of two corrections which
are little discussed. The first issue concerns the effect of the mean
scalar field in the nucleus. In Ref.~\cite{Saito:1995de}
it was shown that the
difference between the radius of a neutron and a proton would grow
with density and it was then claimed that this would lead to a violation
of the
Behrends-Sirlin-Ademollo-Gatto~\cite{Behrends:1960nf,Ademollo:1964sr}
(BSAG) theorem. This result, which
was a consequence of an incorrect application of the constraint equations
in the bag model, is wrong. Indeed, suppose that the $u$ and $d$
masses are equal in a nucleon immersed in an {\it isoscalar} scalar
field. As $m_u$ and $m_d$ move apart the BSAG theorem must apply
whatever the scalar field and one must not obtain a correction
to the Fermi matrix element linear in $m_u - m_d$.
Of course, the coefficient of the term in $(m_u -m_d)^2$ {\it may}
change and we estimate this effect here. It is very small.

On the other hand, when we calculate the change in the Fermi matrix
element in a manner consistent with BSAG, it turns out that at the
physical value of $m_u-m_d$ it is no longer totally negligible
in comparison with the current errors on $V_{ud}$. The reduction
in the overlap of the $u$ and $d$ wave functions turns out to
be of order $2 \times 10^{-5}$. Finally,
we combine this calculation with an earlier estimate of the effect
of pion mass differences carried out using chiral perturbation
theory, in order to obtain a total correction. When applied to the
unitarity test in Eq.~(\ref{eq:unitarity}), the total correction
amounts to about $1.2 \times 10^{-4}$.

As a final note on the possibility of a genuine correction associated
with the nucleon decaying in-medium, we point out that the relatively
small isovector scalar potential (in the QMC model associated with the
$\delta$ meson) expected in nuclei with $N \neq Z$
will serve to effectively increase the value of $m_d - m_u$ by an
amount $V_\delta$. The latter may be as large as 4 or 5 MeV in a
heavy nucleus and this could produce a decrease in the apparent value
of $V_{ud}$ as large as $2 \times 10^{-5}$, depending on the
neutron-proton asymmetry of the particular nucleus. While for
the present this is below the level at which it could be detected
it is a systematic error which may need to be accounted for in the
near future.

\section{Explicit calculation}
In terms of the quark field isodoublet $\psi=(\psi_{u},\psi_{d})$,
the Fermi operator for $\beta$ decay is \[
{\cal F}^{\pm}=\int d\vec{r}\psi^{\dagger}\tau^{\pm}\psi\]
It is related to the generators $(T^{\alpha},\alpha=1,2,3)$ of the
isospin transformations
\[
T^{\alpha}=\int d\vec{r}\psi^{\dagger}\frac{\tau^{\alpha}}{2}\psi,\]
which are conserved by strong interactions in the isospin symmetry
limit. In the following we assume that the quark mass
difference $\delta m=m_{d}-m_{u}$
is the only source of isospin breaking. So we write the exact Hamiltonian
of the strong interaction as
\[H=H_{0}+v
=H_{0}+\frac{\delta m}{2}\int d\vec{r}\bar{\psi}\tau^{3}\psi , \]
where $H_{0}$ is the isospin symmetric part: $[H_{0},T^{\alpha}]=0.$
Even in the presence of the symmetry breaking term $v$ the isospin
generators satisfy the current algebra relations
\[
[T^{\alpha},T^{\beta}]=i\epsilon^{\alpha\beta\gamma}T^{\gamma}\]
and combined with the fact that $[H,T^{\alpha}]\sim\delta m$ this
leads to the BSAG theorem for the Fermi matrix element:
\[
F^{+}=<p|{\cal F}^{+}|n>=1+O(\delta m^{2}) \, . \]

We wish to estimate how the overlap defect $\Delta=1-F^{+}$ changes
when the neutron is in the nuclear medium and the
challenge is to do that in a way consistent with BSAG. When we
naively use the bag model to compute $F^{+}$, as
originally done in the QMC
model, we have the problem that the bag radius depends on the flavor
content. As it changes during the $n\to p$ transition, there is a
lack of orthogonality between the eigenmodes of the initial and final
bag and this leads to an overlap defect of order $\delta m.$ Clearly
this is an artefact of the static bag model which, by definition,
cannot handle a radius which changes with time, as is the case in
a $\beta$ decay.

Since this overlap artefact arises from the small mass difference $\delta m$
the correct approch is to start with the isospin symmetry limit
and use the bag model only to compute the deviation from unity.
We then find that this deviation is explicitly of order $(\delta m)^{2}$
-- as it should be. Higher order deviations can also be computed in this
framework but we stop at order $(\delta m)^{2}$, which is quite
sufficient for our purpose.

We denote as $|N_{\alpha},m_{\tau}=\pm1/2>$, the eigenstates of $H_{0}$
with isospin 1/2.
\[
H_{0}|N_{\alpha},m_{\tau}=\pm1/2>=M_{\alpha}|N_{\alpha},m_{\tau}=\pm1/2>\]
The nucleon corresponds to $\alpha=0$. The exact proton and neutron
states are $|N,m_{\tau}>$ such that
\[
H|N,m_{\tau}>=M(m_{\tau})|N,m_{\tau}>\]
Let $P$ be the projector onto $|N_{0},m_{\tau}=\pm1/2>$ and $Q=1-P.$
We have
\[
|N,m_{\tau}>=\sqrt{Z}\left(1+\frac{Q}{M_{0}-H_{0}}v+\dots\right)|N_{0},m_{\tau}>\]
with
\[
Z=1-<N_{0}|v\frac{Q}{(M_{0}-H_{0})^{2}}v|N_{0}>+O(v^{3})\]
independent of $m_{\tau}.$

Using ${\cal F}^{+}|N_{0},-1/2>=|N_{0},1/2>$ we have, up to $O(v^{3})$
contributions:
\begin{eqnarray*}
F^{+} & = & <N,\frac{1}{2}|{\cal F}^{+}|N,-\frac{1}{2}>\\
& = & Z+\sum_{\alpha\neq0}
<N_{0},\frac{1}{2}|v\frac{1}{(M_{0}-M_{\alpha})^{2}}
|N_{\alpha},\frac{1}{2}><N_{\alpha},-\frac{1}{2}|v|N_{0},-\frac{1}{2}>
\end{eqnarray*}
The terms linear in $v$ vanish because ${\cal F}^{+}$ commutes with
$H_{0}$ and hence with $Q.$ From the isospin structure of the pertubation
we have $<N_{\alpha},-\frac{1}{2}|v|N_{0},-\frac{1}{2}>
= -<N_{\alpha},\frac{1}{2}|v|N_{0},\frac{1}{2}>$.
So the contribution from the excited states adds to that in $Z$
and we end with
\[
F^{+}=1-2<N_{0}|v\frac{Q}{(M_{0}-H_{0})^{2}}v|N_{0}>\]
Since the overlap defect, $1-F^{+}$,
is explicitly of order $(\delta m)^2$,
we can use the (isospin symmetric) bag to finish the calculation.
If we denote as $\omega_{\alpha}$ the energies
with the same quantum numbers
as the nucleon we get
\[
F^{+}=1-\frac{3\delta m^{2}}{2}\sum_{\alpha\neq0}\left(\frac{<\alpha|\gamma_{0}|0>}{\omega_{\alpha}-\omega_{0}}\right)^{2}\]
with
\[
<\alpha|\gamma_{0}|0>=\int^{R}d\vec{r}\,\phi_{\alpha}^{\dagger}
\gamma_{0}\phi_{0} \, , \]
where $\phi_{\alpha}$ are the normalized eigenmodes of the cavity
of radius $R$ with quark mass $m^{*}=\bar{m}-g_{\sigma}^{q}\sigma$
and $\sigma$ is the mean scalar field in the medium, as calculated 
for example in the QMC model~\cite{Guichon:1987jp,Guichon:1995ue,Saito:2005rv}.
In practice we can set $\bar{m}=0.$ Note that the nuclear vector
fields simply shift the energy scale and cannot change $F^{+}.$

\begin{figure}
\includegraphics[clip,width=10cm]{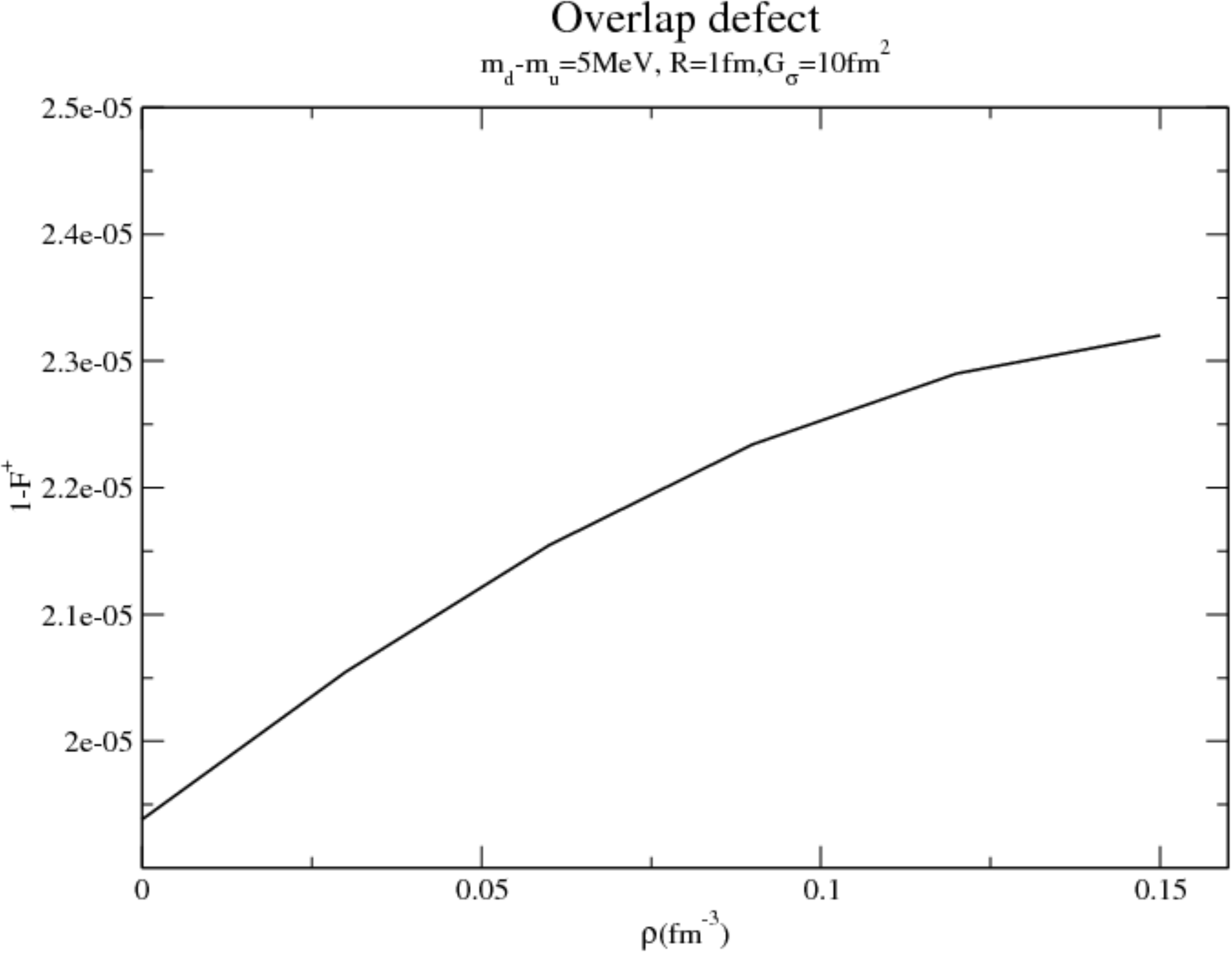}
\caption{
Deviation below unity of the vector matrix element
for the transition neutron to proton, as a function of the
density of isoscalar nuclear matter.}
\label{fig:1}
\end{figure}

If we denote by $\Omega_{\alpha}/R$ the
energy (positive or negative)
of the mode $\alpha$, the eigenmode is
\[
\phi_{\alpha}=\left(\begin{array}{c}
f_{\alpha}(r)\\
i\vec{\sigma}.\hat{r}g_{\alpha}(r)\end{array}\right)\frac{\chi}{\sqrt{4\pi}}\]
with
\begin{eqnarray*}
&&f_{\alpha}(r)  ={\cal N}_{\alpha} \frac{1}{r}\sin\left[\sqrt{\Omega_{\alpha}^{2}-(m^{*}R)^{2}}\frac{r}{R}\right]\\
&&g_{\alpha}(r)  ={\cal N}_{\alpha}
\frac{R}{(\Omega_{\alpha}+m^{*}R)r}\\
&&\left(\frac{1}{r}
\sin\left[\sqrt{\Omega_{\alpha}^{2}-(m^{*}R)^{2}}
\frac{r}{R}\right]-\frac{\sqrt{\Omega_{\alpha}^{2}-(m^{*}R)^{2}}}{R}
\cos\left[\sqrt{\Omega_{\alpha}^{2}-(m^{*}R)^{2}}
\frac{r}{R}\right]\right)
\end{eqnarray*}
and ${\cal N}$ the normalization constant such that
\[
\int_{0}^{R}r^{2}dr\left(f_{\alpha}^{2}+g_{\alpha}^{2}\right)=1 \, .
\]
The energy is determined by the boundary condition $f(R)=g(R)$,
that is
\begin{eqnarray*}
(\Omega_{\alpha}+m^{*}R)\sin\left[\sqrt{\Omega_{\alpha}^{2}-(m^{*}R)^{2}}
\right]&=&\sin\left[\sqrt{\Omega_{\alpha}^{2}-(m^{*}R)^{2}}\right]\\
&&-\sqrt{\Omega_{\alpha}^{2}-(m^{*}R)^{2}}
\cos\left[\sqrt{\Omega_{\alpha}^{2}-(m^{*}R)^{2}}\right] \, .
\end{eqnarray*}
In the following we assume that $m^{*}R$ is larger than the critical
value $-3/2$ at which $\Omega_{0}=3/2.$ This is not a restriction
in normal nuclei. The scalar matrix element is then
\[
<\alpha|\gamma_{0}|0>=\int_{0}^{R}r^{2}\left(f_{\alpha}f_{0}-g_{\alpha}g_{0}
\right)dr \, .
\]
We show the resultant overlap defect as a function of density for
nuclear parameters~\cite{Guichon:1995ue}
in Fig.~\ref{fig:1}. The $u-d$ mass difference has been set to
5 MeV, which is appropriate to the MIT bag~\cite{Bickerstaff:1981ut}.
The overlap defect changes by only a few
times $10^{-6}$ as the density varies from zero to
nuclear matter density.

\section{Discussion}
Although the variation of the Fermi matrix element with density
is small, we note from Fig.~\ref{fig:1} that the intercept
at zero density is around $1.94 \times 10^{-5}$, which is only
a factor of 10 smaller than the current error quoted for $V_{ud}$.
While such a deviation from unity for the hadronic form factor is
carefully handled in the $s \rightarrow u$ transition, it seems
to receive little attention in the literature in the
$d \rightarrow u$ transition. Indeed, we found only the discussion
by Kaiser\cite{Kaiser:2001yc},
in the framework of chiral perturbation theory.
That work reported a total reduction from the mass difference
between charged and neutral pions of about $4 \times 10^{-5}$. The
additional contribution from the difference between the $\pi^0 nn$ and
$\pi^0 pp$ coupling constants,
which is of order 0.4\%~\cite{Thomas:1981mj},
produces a negligible defect, well below $10^{-6}$.
The pionic correction of Kaiser and the
correction from the valence quark
overlap calculated here are independent and hence the total
correction is $\delta g_V = 6 \times 10^{-5}$.
Thus one should increase the value of $V_{ud}$ deduced from the
analysis of super-allowed Fermi $\beta$-decay by this amount.
As noted earlier, this amounts to an increase in the value
of $|V_{ud}|^2 \, + \, |V_{us}|^2 \, + \, |V_{ub}|^2$ by
around $1.2 \times 10^{-4}$, which is only a factor of 5 below
the error quoted in Eq.~(\ref{eq:unitarity}).

While the absence of a sizeable correction to $g_V$ in-medium is
reassuring, we note that there is a genuine correction which has
so far been ignored, involving the isovector mean scalar field.
This is often ignored because the coupling is so much smaller than
that for the isoscalar case, but if the mean isovector scalar field felt
by the quark is $\frac{\tau_z}{2} V_\delta$, the isospin breaking term
will be proportional to $(m_d - m_u + V_\delta)^2$. This could
potentially double the deviation associated with the quark mass
difference, depending on the neutron excess and structure of the
particular nucleus under consideration. It will be interesting
to explore this effect quantitatively in future as well as to
investigate the model dependence of isospin breaking term
calculated here within the MIT bag model.

\section*{Acknowledgements}
This work was suppported by the CEA (PAMG) and by an Australian Laureate
Fellowship and the University of Adelaide (AWT).

\end{document}